\documentclass[preprint,aps,showpacs]{revtex4}
\usepackage{epsfig}

\newcommand{\Be}{\begin{equation}}
\newcommand{\Ee}{\end{equation}}
\newcommand{\Bea}{\begin{eqnarray}}
\newcommand{\Eea}{\end{eqnarray}}

\begin{document}

\title{Charmed Mesons Have No Discernable Color-Coulomb Attraction}
\author{T.\ Goldman}\email{tgoldman@lanl.gov}
\affiliation{Theoretical Division, MS-B283, 
Los Alamos National Laboratory, Los Alamos, NM 87545}
\author{Richard R.\ Silbar}\email{silbar@lanl.gov} 
\affiliation{Theoretical Division, MS-B283, 
Los Alamos National Laboratory, Los Alamos, NM 87545}

\begin{flushright}
\today\\
{LA-UR-11-11069}\\
{arXiv:yymm.nnnn}\\
\end{flushright}

\begin{abstract}
Starting with a confining linear Lorentz scalar potential $V_s$ and a Lorentz vector 
potential $V_v$ which is also linear but has in addition a color-Coulomb attraction 
piece, $-\alpha_s/r$,
we solve the Dirac equation for the ground-state $c$ and $u$ quark wave functions.
Then, convolving $V_v$ with the $u$-quark density, we find that the Coulomb attraction
mostly disappears, making an essentially linear $\bar{V}_v$ for the $c$-quark.
A similar convolution using the $c$--quark density also leads to an essentially linear 
$\tilde{V}_v$ for the $u$-quark.
For bound $\bar{c}$-$c$ charmonia, where one must solve using a reduced mass for the
$c$-quarks, we also find an essentially linear $\widehat{V}_v$.
Thus, the relativistic quark model describes how the charmed-meson mass spectrum
avoids the need for a color-Coulomb attraction. 
\end{abstract} 


\maketitle

\section{Introduction}

Non-relativistic potential models for charmonium states~\cite{Rosner} have done 
remarkably well at describing these states with a simple linear confining 
potential (plus spin-spin and spin-orbit terms), absent any evidence for a short 
distance color-Coulomb contribution. The last is somewhat surprising as, early on, the high 
mass of the charm quark suggested that the color-Coulomb region might be discernible, 
at least in the wave functions, if not the eigenenergies. 

Although a Dyson-Schwinger approach would be even more appropriate~\cite{ANL}, we report here 
on the results of attempting a self-consistent potential approach using the Dirac equation 
with both color Lorentz vector and confining color Lorentz scalar potentials designed to 
match all available data. Our principle result is that an iterative self-consistency requirement 
on these potentials leads to a color vector potential that is virtually indistinguishable from 
linear over the full range of interest. 

We take our viewpoint from the relativistic approach to the Hydrogen atom: the Dirac 
equation is used for a reduced mass electron in a potential determined by the total 
charge interior to the radial point under consideration. This approach has been  
studied intensively by, for example, Friar and Negele~\cite{FN}. The closest analogy, 
then, is to the case of D-mesons, with one light quark and a charm quark in the analogous 
role to that of the proton in Hydrogen. 

As the light quark mass is negligible on the scale of interest (before we approach 
refinements of electromagnetic accuracy), there is no discernible reduced mass 
effect to consider. Furthermore, in Hydrogen, the charge distribution within the proton 
(or nucleus in more massive atoms) smears out the Coulomb divergence at zero 
separation. Here, the charm quark is the color Coulomb source but has no intrinsic 
internal structure. However, unlike the electromagnetic case, the virtual emission and 
reabsorption of gluons, produces fluctuations in the color source location that are not 
negligible even to leading order. Our approach is to take, as a first approximation, the 
charge distribution (rms size) of the D-meson as setting a relevant scale for these 
fluctuations in smearing the color Coulomb divergence. Although it may not be precise, 
given that the rms size is comparable to the inverse of the QCD mass scale, we consider this to be a 
very reasonable starting point. We use this scale to build a smooth truncation of the 
color Coulomb potential with a quadratically flat ``bottom" and with slope equal to that 
of the $-\alpha_{s}/r$ color Coulomb potential at the matching point. 

We have checked that such a potential produces a reasonable representation of the 
charmonium spectrum ($\sim \pm 50$ MeV).  We then convolve the light quark wavefunction 
with the color Coulomb potential in the manner of Friar and Negele \cite{FN} to define 
the potential that the charm quark is subject to in the presence of the light quark. Finally, 
to check for self-consistency, we again convolve this charm quark wavefunction with the 
color Coulomb potential and observe its effect on the light quark. We find a consistent, 
almost precisely linear, effective vector potential radially out all the way to the region 
where linear confining potentials are necessary for consistency with data. 

We then repeated this sequence for charmonium states where reduced mass effects are 
no longer negligible. We find a very similar result and conclude that there is indeed a 
single, consistent, approximately linear color vector potential that reasonably describes 
all of these states, despite the complications one might expect from the detailed issues 
described above. Thus, there is no remaining evidence of the color Coulomb potential 
despite the relatively large mass of the charm quark. The bottom quark may another matter.

\section{Our Original Set of Potentials}

After some experimentation (and with some prejudices) we came to consider the 
charmed $D$-mesons [$\bar{c}$-$u$] and (bound) charmonia [$\bar{c}$-$c$] as resulting 
from scalar and vector potentials,
$V_s(x)$ and $V_v(x)$, like those shown in Fig.\ 1.

\begin{figure}[t] 
\includegraphics[width=0.6\textwidth, height=0.375\textwidth, angle=0]{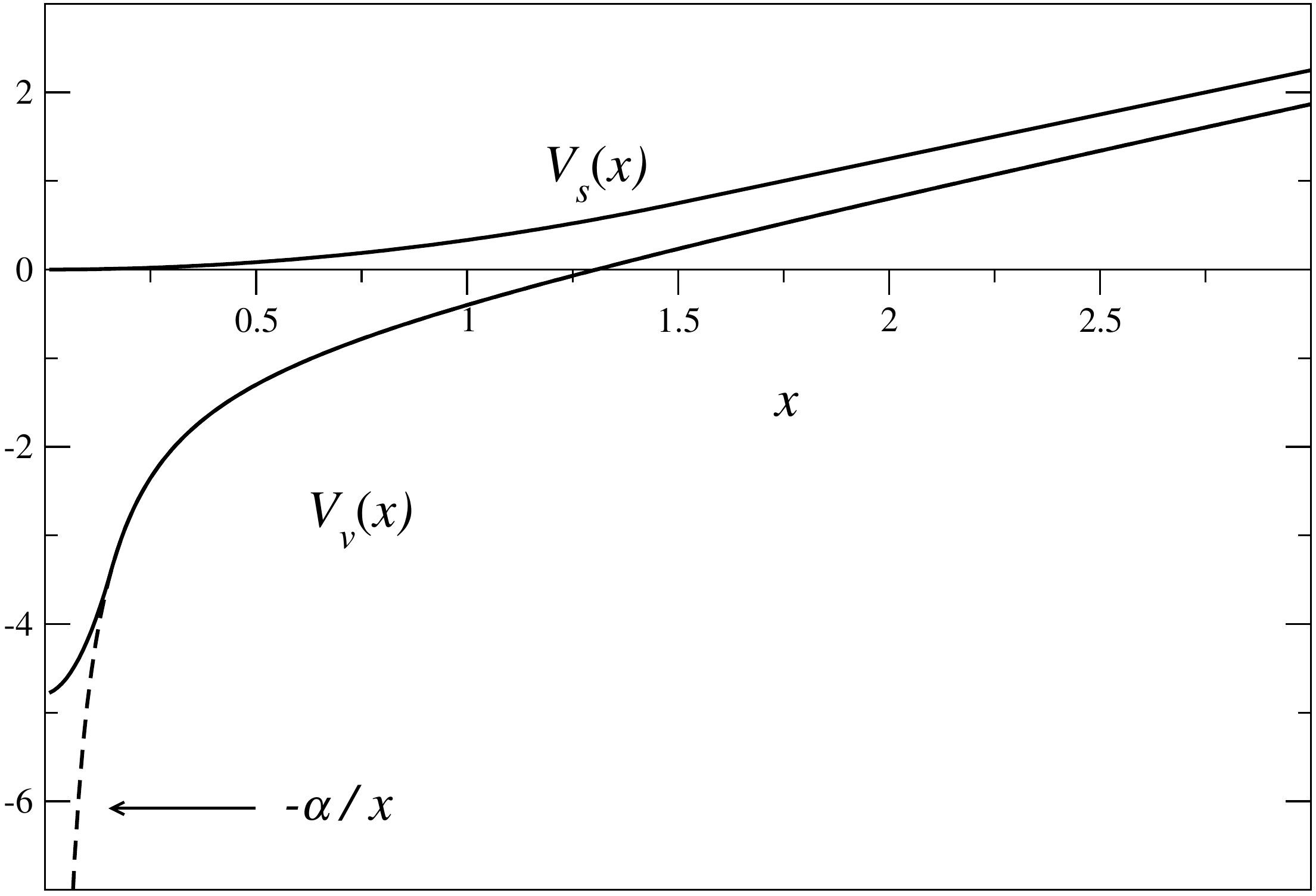}
\caption{The scalar and vector potentials $V_s(x)$ and $V_v(x)$ 
as functions of a dimensionless radial coordinate, $x$.}
\label{fig:VsVv0}
\end{figure}

These potentials are dimensionless functions of a dimensionless radial coordinate $x=\kappa r$,
where $\kappa^2$ is taken here to be 0.9 GeV/fm \cite{GMSS}, i.e., $\kappa$ = 2.136 fm$^{-1}$.
The asymptotically linear slopes of $V_s$ and $V_v$ are the same, in accordance with the
small spin-orbit splitting in the baryon spectrum.\cite{Isgur, PGG}
The $V_s$ is quadratic out to $x = 1.5$ after which it is strictly linear:
\Be 
		V_s(x) = \left\{ \begin{array}{ll}
		x^2/(2 x_s) & \quad\mbox{if $x < x_s$,} \\
		x - x_s/2   & \quad\mbox{otherwise} \ ,
		\end{array}
		\right. \label{eq:quadVs}
\Ee  
where the parameter $x_s$ is, for us, fixed at 1.5.
The flatness near $x=0$ is to preserve chiral symmetry at short distances.

The $V_v$ has, in addition to the linear confinement, a color-Coulomb contribution,
\Be 
	V_v(x) = -\frac{\alpha_s}{x} + x - x_v \ , \label{eq:VvCoul}
\Ee
as shown by the dashed part in Fig.\ 1.
(For this figure, we used $x_v= 1.0$ and $\alpha = 0.4$.)
However, this is the potential seen by, say, the light $u$-quark in the field of 
the heavy $\bar c$-quark, which is itself moving about somewhat in the field of the $u$-quark.
Thus it is reasonable to moderate the singularity at $x=0$.
We did this simply by altering the potential to
\Be 
	V_v(x) = \left\{ \begin{array}{ll}
		\alpha_s \; (x^2 - 3 x_D^{\;\; 2})\,/\,2 x_D^{\;\; 3} \ + \ x - x_v, 
			& \quad\mbox{if $x < x_D$,} \\
		-\alpha_s /x \ + \ x - x_v, & \quad\mbox{otherwise} \ ,
		\end{array}
		\right. \label{eq:roundedVv}
\Ee
assuming the smoothing to be about the size of the (electric) charge radius ($R_D$) of 
the $D(1869)$ meson.
Here $x_D = 0.16$ is a reasonable guess (corresponding to $R_D \sim 0.3$ fm), 

It turns out that solving the radial Dirac equations \cite{EJP} for a charmed quark mass of 
$m_c = 1.550$ GeV for the potentials as shown in Fig.\ 1 leads to a reasonably accurate 
description of the masses of the (bound) charmonia $\bar{c}\; c$ states -- 
$\eta_c$, $J/\psi$, $\eta_c'$, $\psi'$, $\chi_0$, and $\chi_2$.\ \cite{PDG}
(For these charmonia calculations, the mass in the coupled differential equations 
must be the reduced mass, $m_c/2$.)
Similar calculations for $\bar{c}\;u$, however, do {\it not} accurately reproduce the 
charmed $D$-meson masses.

\section{Convolving the Coulomb potential for the $\bar c$-quark}

A more consistent way of moderating the Coulomb singularity is first to solve for the 
light $u$-quark $1s$ ground-state wave functions \cite{EJP} for the potentials  
$V_s(x)$ [Eq.\ (\ref{eq:quadVs})] and 
$V_v$ [Eq.\ (\ref{eq:roundedVv}), with $x_v=1.0$ and $\alpha_s = 0.4$].
Then, the vector potential that the $\bar c$-quark should be subject to is the (unrounded) $V_v$ given by
Eq.\ (\ref{eq:VvCoul}) modulated by the density of that $u$-quark.
That is, following Friar and Negele's discussion of muonic atoms \cite{FN},
the Coulomb potential should be convolved with the local ``charge'' density defined by the $u$-quark Dirac wave function.
That, together with the linear contribution, gives a new vector potential,
\Be
	\bar{V}_v(x) = Q_{\rm in}(x)/x + Q_{\rm out}(x) + x - x_v \ , \label{eq:newVv}
\Ee
where
\Bea
	Q_{\rm in}(x)/x &=& -\frac{\alpha_s}{x} \int_0^x x'^{\;2} dx'\; 
		\psi_{u,1s}^\dagger(x') \psi_{u,1s}(x') 
		= -\frac{\alpha_s}{x} \int_0^x x'^{\;2} dx'\; [\psi_{a}^2(x') + \psi_{b}^2(x')] \ , 
		\nonumber \\
	Q_{\rm out}(x) &=& -\alpha_s \int_x^\infty x' dx' \; \psi_{u,1s}^\dagger(x') \psi_{u,1s}(x') 
		= -\alpha_s \int_x^\infty x' dx' \; [\psi_{a}^2(x') + \psi_{b}^2(x')] \ ,
		\label{eq:QandQpr}
\Eea
where $\psi_a(x)$ is the (real) upper component and $\psi_b(x)$ the lower component.
Near $x=0$,
\Be 
	Q_{\rm in}(x)/x \approx \frac{|\psi(0)|^2}{x} \int_0^x x'^{\;2} dx' =
		\frac{|\psi(0)|^2 \; x^2}{3} \label{eq:Qbyxnear0}
\Ee
and it never gets more negative than about $-0.2$ before it increases again toward zero
like $-\alpha_s/r$.
As for $Q_{\rm out}(x)$, since the $1s$ upper component radial wave function is 
well-approximated as a Gaussian \cite{GMSS, Critch}, its integral gives, approximately, 
$-\alpha_s$ times a (narrower) Gaussian.

\begin{figure}[t] 
\includegraphics[width=0.6\textwidth, height=0.375\textwidth, angle=0]{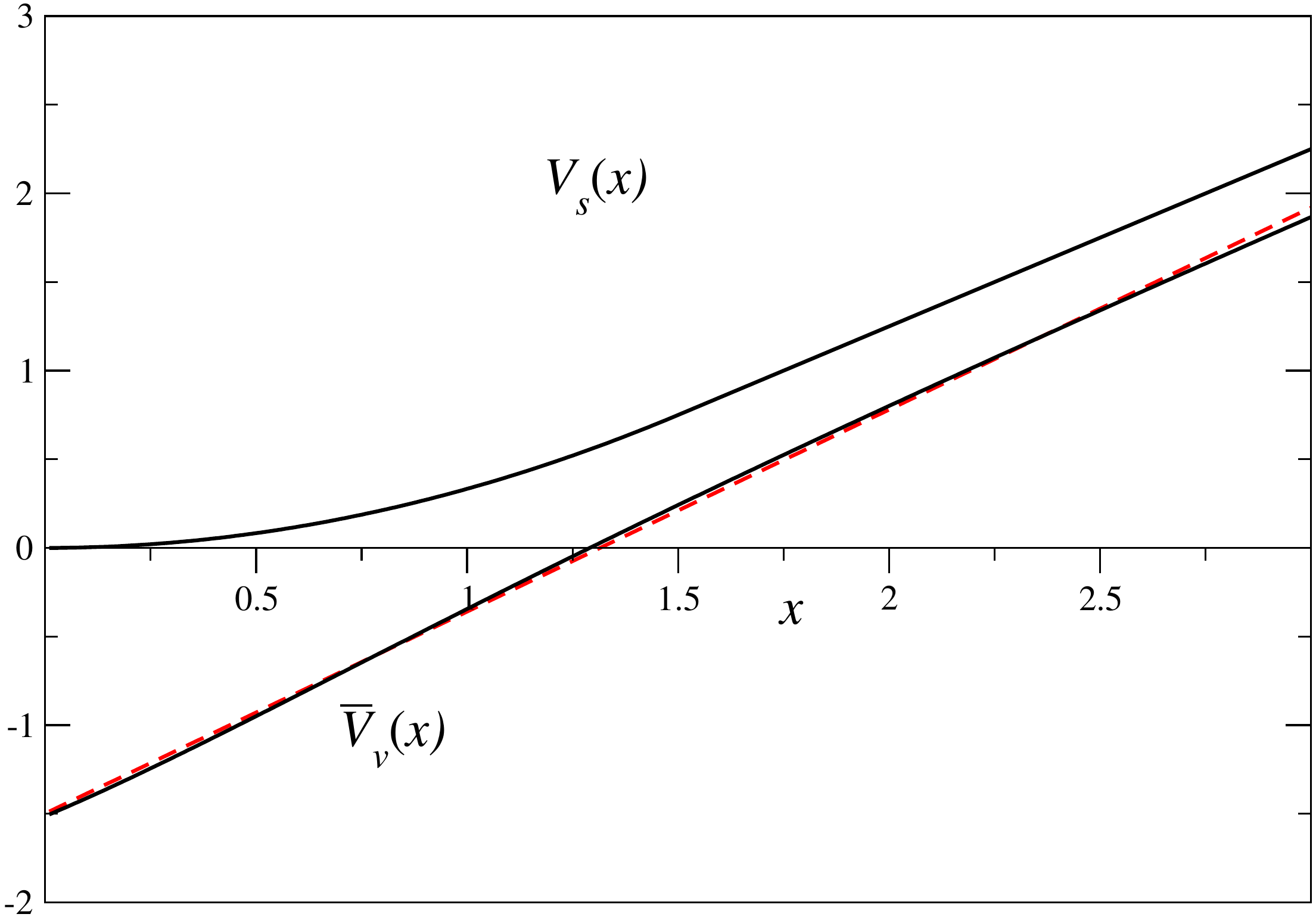}
\caption{(Color online) The scalar potential $V_s(x)$ and the convolved vector potential $\bar{V}_v(x)$ 
for the $\bar c$-quark as functions of $x$.}
\label{fig:VsVv1}
\end{figure}

A plot of $\bar{V}_v(x)$ calculated from the integrals of Eq.\ (\ref{eq:QandQpr}) is given 
in Fig.\ 2.
Despite its appearance, it is {\it not strictly} a straight line -- 
there is some small curvature in the plot below $x=1$. Nonetheless, we consider
the high accuracy of a linear approximation to be rather surprising, 
as we were expecting only a minor change in the effective value of $x_D$.
The dashed line in Fig.\ 2 is a linear fit to $\bar{V}_v(x)$, $a_0 x + a_1$, with slope 
$a_0 = 1.139$ and displacement $a_1 = -1.498$.

\section{How $\psi_c$ changes with the new potential}

If one solves for the $1s$ state of the $\bar c$-quark for $m_c = 1.550$ GeV and the 
original potentials of Eq.\ (\ref{eq:quadVs}) and Eq.\ (\ref{eq:roundedVv}), 
one finds the $\bar c$-quark eigenenergy to be $E_c(1s) = 1.339$ GeV.
That is, the energy of the $c$-quark for these potentials is some 200 MeV 
less than its mass.
The upper and lower $1s$ radial wave functions are displayed in Fig.\ 3 as dashed
curves.
\begin{figure}[t] 
\includegraphics[width=0.6\textwidth, height=0.375\textwidth, angle=0]{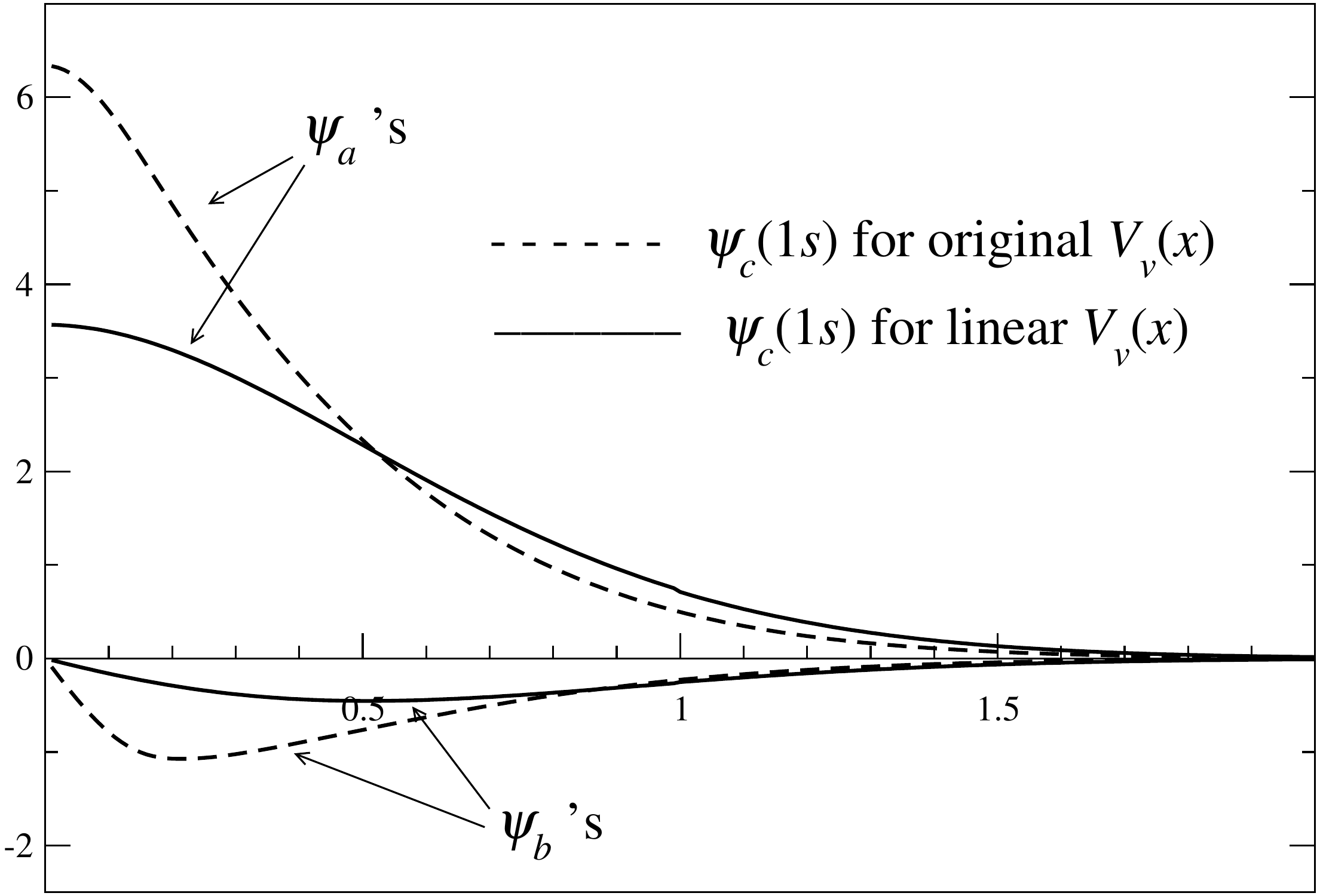}
\caption{Comparing the $\bar c$-quark radial wave functions for the original and 
convoluted potentials.}
\label{fig:psiwfns}
\end{figure}
The rise of $\psi_a$ at the origin is reminiscent of the non-relativistic ground state
wave function for a pure Coulomb potential, 
which is a wave function that is a decaying exponential like $e^{-\alpha x}$.

However, solving for $E_c(1s)$ with $V_s(x)$ and the convolved $\bar{V}_v(x)$,
we find $E_c(1s) = 1.534$ GeV, now considerably higher in energy because of the 
missing Coulomb well, although still less than $m_c$.
The upper and lower component wave functions, shown as the solid curves in Fig.\ 3,  
are somewhat broader than those found for the original potentials.
(In both cases, however, the $\bar c$-quark wave functions are not as broad as those
for the $u$-quark.)

Thus, while the mass of a $\bar{c} u$ $D$-meson, which depends strongly on $E_c(1s)$,
is not very sensitive to the difference between the original and convolved potentials,
the wave functions are quite different.  
We therefore expect that quantities such as transition strengths, which are more
dependent on the details of the wave functions, will be more potential-dependent, as usual.

\section{Convolving for the $u$-quark potential}

The singular Coulomb piece of the vector potential seen by the $u$-quark is 
also smeared by the motion of the somewhat more confined, slower moving $\bar c$-quark.
With formulae like Eqs.\ (\ref{eq:newVv}) and (\ref{eq:QandQpr}), but with the density 
provided by $|\psi_c(1s)|^2$, we again found this second convolution is also very close to
linear, as shown in Fig.\ 4.
\begin{figure}[t] 
\includegraphics[width=0.6\textwidth, height=0.375\textwidth, angle=0]{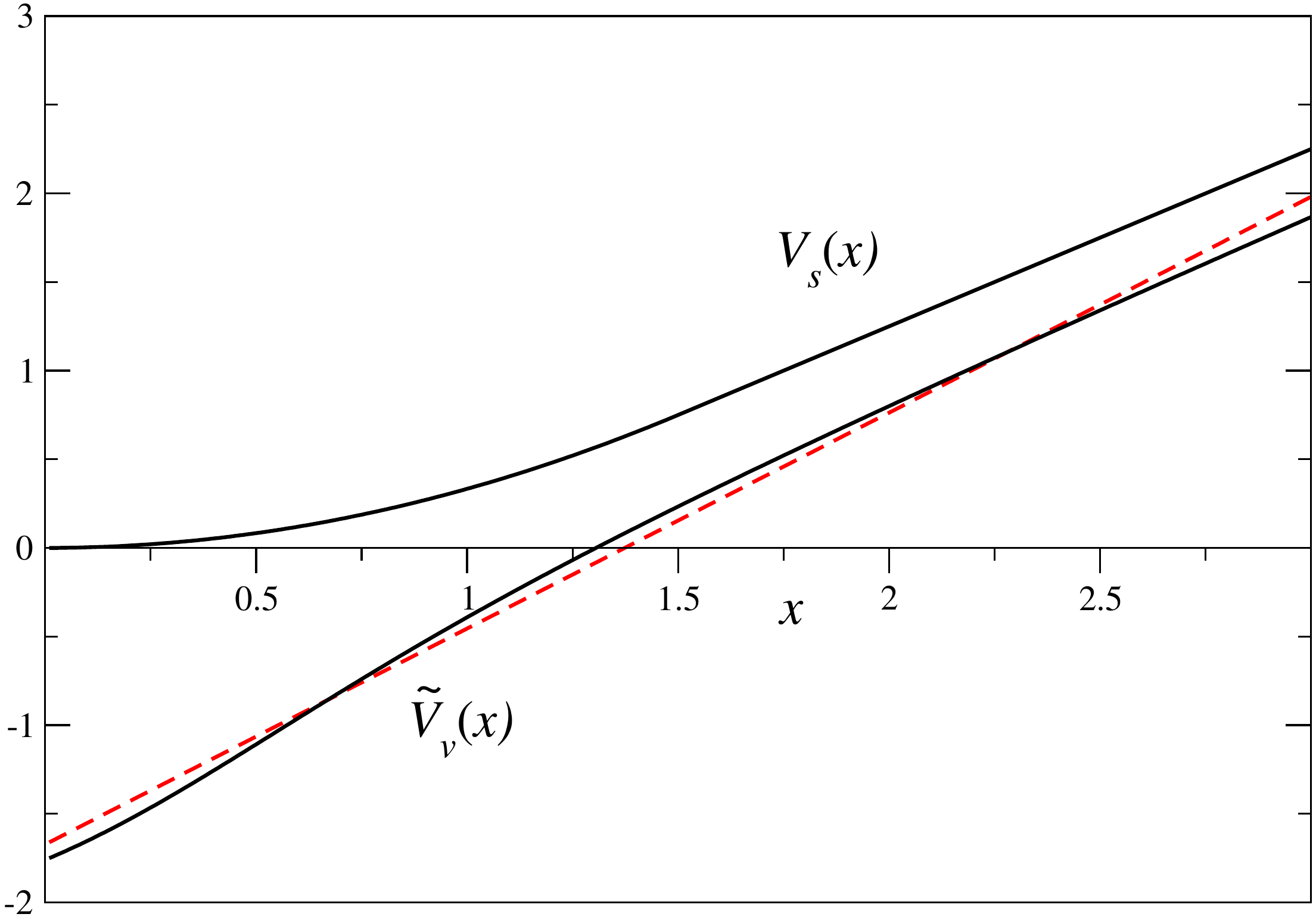}
\caption{(Color online) The scalar potential $V_s(x)$ and the convolved vector potential 
$\tilde{V}_v(x)$ for the $u$-quark as functions of $x$.}
\label{fig:VsVv2}
\end{figure}
In making this plot we used the $\psi_c(1s)$ found from using $\bar{V}_v$.
Again, the dashed line is a linear fit to $\tilde{V}_v(x)$ with slope $a_0 = 1.218$
and displacement $a_1 = -1.673$.

The $\tilde{V}_v$ is slightly more curved and a bit deeper than the $\bar{V}_v$ shown in Fig.\ 2,
but is essentially the same nearly-linear potential as that which affects the $\bar c$-quark.
Although the linear fit parameters are slightly different, one could reasonably assume the 
same linear potential for both quarks. 

\section{Convolving for charmonium}

Similarly, we investigated the smearing of the Coulomb potential for the charmonia 
[$\bar c$-$c$] states.
In this case we must solve for the {\it reduced} mass, which is $m_c/2 = 0.775$ GeV, 
but otherwise the calculation proceeds much as above.
First we find the wave functions for the $c$-quark for the original potentials,
$V_s(x)$ [Eq.\ (\ref{eq:quadVs})] and $V_v$ [Eq.\ (\ref{eq:roundedVv})].
Then, convolving the singular Coulomb potential with $|\psi_{c,{\rm reduced}}|^2$
as in Eqs.\ (\ref{eq:newVv}) and (\ref{eq:QandQpr})
we obtain the plot of, now, $\widehat{V}_v$ shown in Fig. 5. 
The dashed line is a linear fit to $\tilde{V}_v(x)$ with slope $a_0 = 1.225$
and displacement $a_1 = -1.688$. Note that the linear fit parameters are 
extremely close to the values that we found for the (effective) potential for 
the $u$-quark.
\begin{figure}[t] 
\includegraphics[width=0.6\textwidth, height=0.375\textwidth, angle=0]{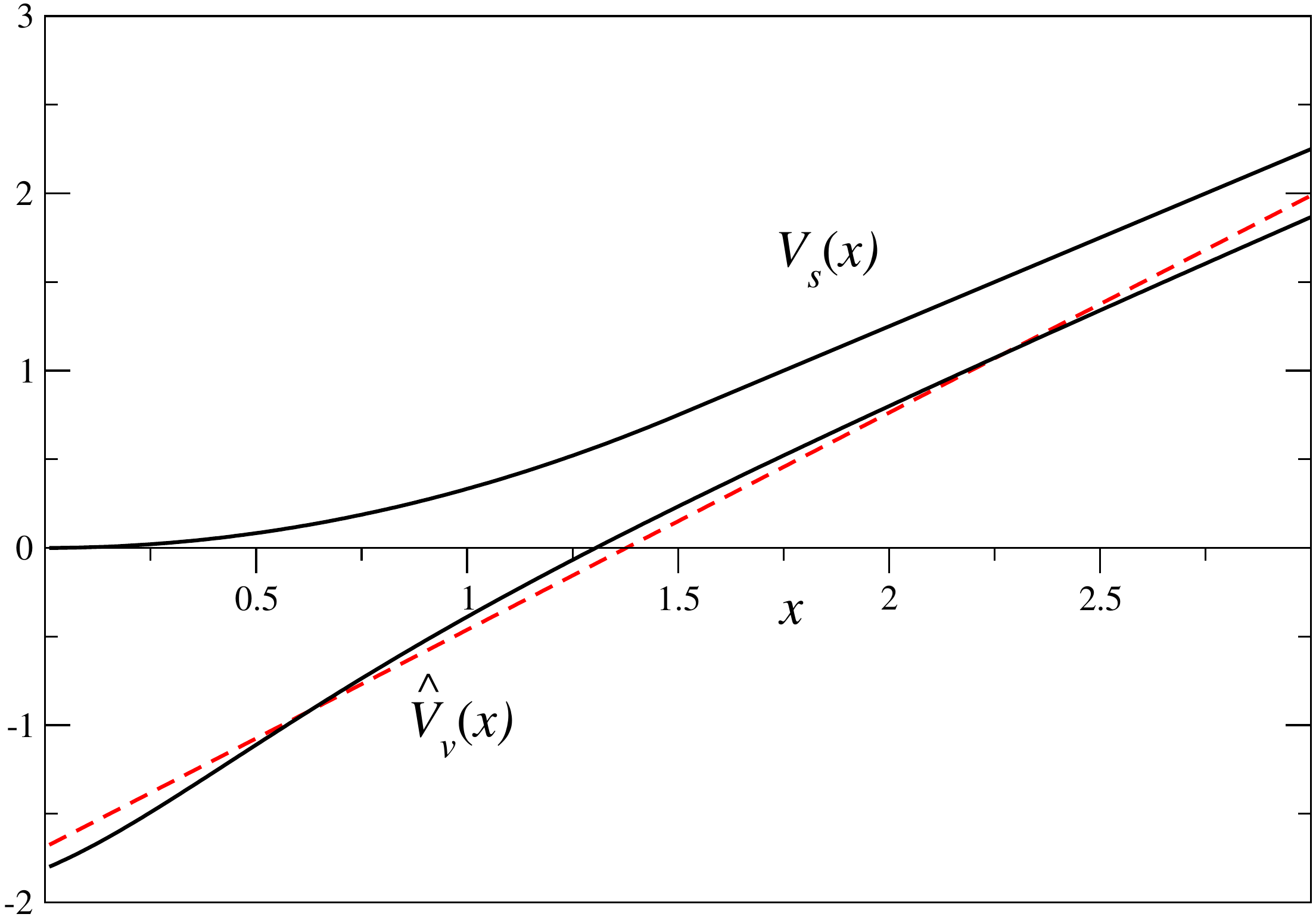}
\caption{The scalar potential $V_s(x)$ and the convolved vector potential 
$\widehat{V}_v(x)$ for charmonia states as functions of $x$.}
\label{fig:VsVv3}
\end{figure}

\section{Discussion}

These results are consistent with the good spectral results found in non- and relativistic 
models for the charmonium spectrum using linear confining potentials. It is unclear 
whether the slight differences in the effective linear potentials merit the complications 
of a relativistic approach to the calculation of spectra until high accuracies become 
necessary. Our convolution approach suggests that the good spectral results with a 
non-relativistic, linear potential are due to the still relatively light mass of the charm quark. 
This in turn invites the question as to whether this status can still hold for the bottom 
quark. Absent detailed comparisons with transition rates for charmonium and D-meson 
decays, which one expects to be more sensitive to wave function details than are spectra, 
the color Coulomb contribution to the effective potential for quark binding remains 
undetermined from charmonium data alone. We intend to turn next to bottom quark states 
to examine whether the spectra there can provide a definitive determination of this issue. 

\section{Acknowledgments}

We thank James Friar for an illuminating conversation which led us to the convolutions
in Eq.\ (\ref{eq:QandQpr}). This work was carried out in part under the auspices of the 
National Nuclear Security Administration of the U.S. Department of Energy at Los Alamos 
National Laboratory under Contract No. DE-AC52-06NA25396.

\end{document}